\documentclass[sigplan]{acmart}

\pdfoutput=1

\usepackage[normalem]{ulem}
\usepackage[T1]{fontenc}
\usepackage{algpseudocode}
\usepackage{enumitem}
\usepackage{listings}
\usepackage{courier}
\usepackage{amsmath}
\usepackage{dblfloatfix}
\usepackage{graphicx}
\usepackage{tabularx}
\usepackage{multirow}
\usepackage{bigdelim}
\usepackage{booktabs}
\usepackage{wrapfig}
\usepackage{mwe}
\usepackage{balance}

\copyrightyear{2020}
\acmYear{2020}
\setcopyright{acmlicensed}
\acmConference[CC '20]{Proceedings of the 29th International Conference on Compiler Construction}{February 22--23, 2020}{San Diego, CA, USA}
\acmBooktitle{Proceedings of the 29th International Conference on Compiler Construction (CC '20), February 22--23, 2020, San Diego, CA, USA}
\acmPrice{15.00}
\acmDOI{10.1145/3377555.3377893}
\acmISBN{978-1-4503-7120-9/20/02}

\author{Philip Ginsbach}
\affiliation{
  \department{School of Informatics}
  \institution{The University of Edinburgh}
  \country{United Kingdom}}
\email{philip.ginsbach@ed.ac.uk}

\author{Bruce Collie}
\affiliation{
  \department{School of Informatics}
  \institution{The University of Edinburgh}
  \country{United Kingdom}}
\email{bruce.collie@ed.ac.uk}

\author{Michael F.\ P.\ O'Boyle}
\affiliation{
  \department{School of Informatics}
  \institution{The University of Edinburgh}
  \country{United Kingdom}}
\email{mob@ed.ac.uk}

\definecolor{sh_keyword1}{rgb}{0.5, 0, 0} 
\definecolor{sh_keyword2}{rgb}{0, 0, 0.5}
\definecolor{sh_keyword3}{rgb}{0, 0.5, 0} 

\lstdefinelanguage{LiLAC}
{
morekeywords = { HARNESS, IMPLEMENTS, Marshaling, PersistentVariables,
                 BeforeFirstExecution, AfterLastExecution, CppHeaderFiles,
                 INPUT, OUTPUT, COMPUTATION},
morekeywords = [2]{CudaRead, CudaWrite, spmv_csr, Maximum, cuda, mkl,
                   dotproduct},
morekeywords = [3]{out, in, size},
morecomment  = [s]{"}{"}
}

\lstdefinelanguage{constraints}
{
morekeywords = { Constraint, with, and, or, at, as, End },
morecomment  = [s]{\{}{\}}
}

\begin{document}

\title{Automatically Harnessing Sparse Acceleration}
\date{}

\begin{abstract}
Sparse linear algebra is central to many scientific programs, yet compilers fail
to optimize it well.
High-performance libraries are available, but adoption costs are
significant.
Moreover, libraries tie programs into vendor-specific software and hardware
ecosystems, creating non-portable code.

In this paper, we develop a new approach based on our {\em specification
Language for implementers of Linear Algebra Computations} (LiLAC).
Rather than requiring the application developer to (re)write every program for a
given library, the burden is shifted to a {\em one-off} description by the
library implementer.
The LiLAC-enabled compiler uses this to insert appropriate library routines
without source code changes.

LiLAC provides automatic data marshaling, maintaining state between calls and
minimizing data transfers.
Appropriate places for library insertion are detected in compiler intermediate
representation, independent of source languages.

We evaluated on large-scale scientific applications written in FORTRAN;
standard C/C++ and FORTRAN benchmarks; and C++ graph analytics
kernels.
Across heterogeneous platforms, applications and data sets we show speedups of
1.1$\times$ to over 10$\times$ without user intervention.

\end{abstract}



\maketitle

\section{Introduction}
\label{sec:introduction}
Linear algebra is an important component of many applications and a prime
candidate for hardware acceleration.
While there has been significant compiler effort in accelerating dense algebra
\cite{Grosser2012Polly,Rapaport:2015:SWF:2863697.2864617,
      Nuzman:2011:VSA:2190025.2190062},
there has been less success with sparse codes.
This is largely due to indirect memory access, which challenges compiler
analysis \cite{Mohammadi:2019:SCD:3314221.3314646}.
Sparse-based algorithms are, however, increasingly important as the basis of
graph algorithms and data analytics \cite{Kepner2015GraphsMA}.

We currently see the wide-scale provision of fast sparse libraries
\cite{cusparse,clsparse,mkl,oski}.
They deliver excellent performance, but require significant programmer
intervention and are rarely portable across platforms.
Alternatives, such as the SLinGen/LGen system
\cite{Spampinato:14,Spampinato:18}, provide specialized code generators for
linear algebra, but again require code modification by the programmer and focus
only on dense computations.

Program modification is particularly problematic when the targets are hardware
accelerators that require careful data marshaling.
Such modifications are often program-wide and severely reduce the portability of
the program.
Furthermore, they require a commitment to specific hardware vendors, resulting
in codebases that quickly become obsolete.
In order to mitigate this, many projects have to keep multiple execution paths,
resulting in arcane build systems
and unmaintainable code.
In this time of rapid hardware innovation, such a vendor lock-in is undesirable.
In fact, the difficulty of efficient portable integration is a key impediment to
the wider use of accelerator libraries and hardware.

In this paper, we reexamine how compilers and libraries can be used to achieve
performance without programmer effort.
Highly tuned and platform specific-libraries invariably remain the fastest
implementations available.
However, we show that we can automatically integrate these libraries without
polluting the source code.
This is performed as a compiler transformation step, leaving the original source
code intact and portable.

To achieve this, we develop a new specification language for implementers of libraries,
the {\em specification Language for implementers of Linear Algebra
Computations} (LiLAC).  Using LiLAC, library implementers specify
with a few lines of code, {\em what} a library does and {\em how} it
is invoked.  Our compiler then determines where the library
specification matches user code and automatically transforms it to
utilize the library.
The language has two complementing parts.

\begin{figure*}[t]
\includegraphics[width=\textwidth]{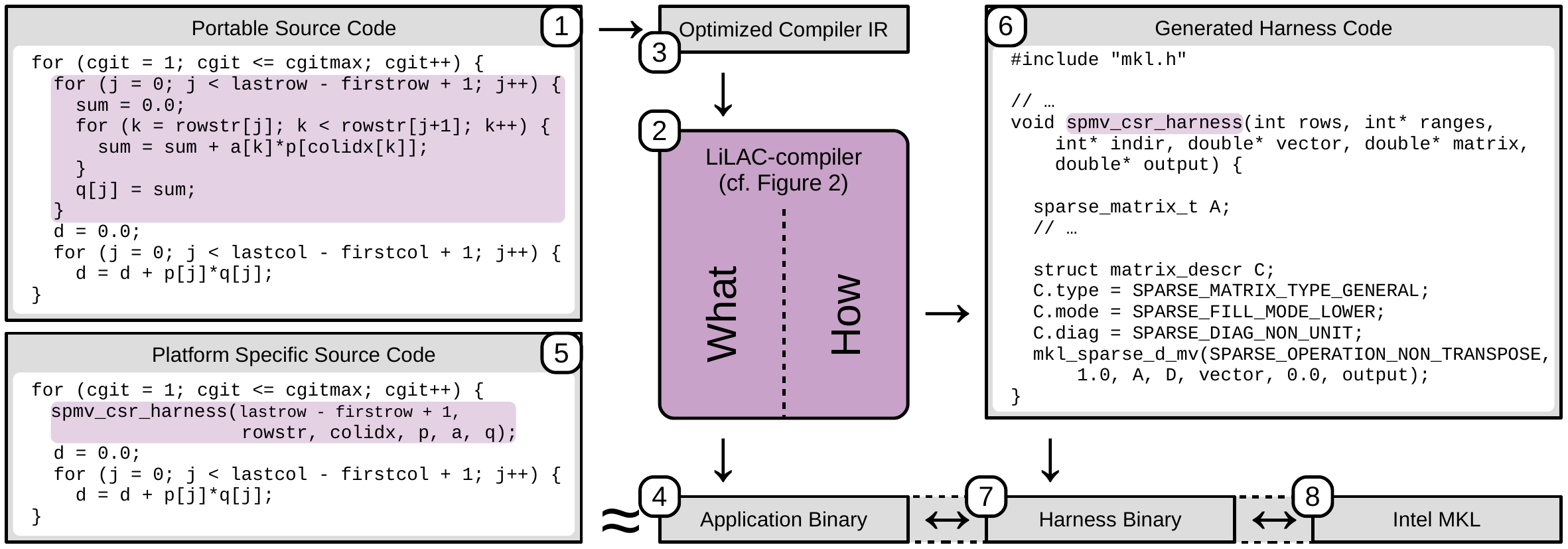}\\[-0.5em]
\caption{LiLAC applied to NPB Conjugate Gradient:
         Code (1) that matches the LiLAC-What specification
         (cf. \autoref{motivationflow}) is replaced by
         calls to a harness (5) during compilation (2), resulting in
         an application binary (6) that corresponds to (hypothetical)
         platform-specific source code (4).
         The harness is generated from the LiLAC-How specification
         (cf. \autoref{motivationflow}) to utilize Intel MKL.
         \parfillskip=0pt}
\label{motivationcode}
\end{figure*}

\noindent
{\bf LiLAC-What} is a high-level language to describe sparse and dense linear
algebra computations.
The LiLAC compiler uses it to detect such functionality
in user applications at compiler intermediate representation level.
It is powerful enough to formulate linear algebra routines, yet remains
independent of compiler internals and is easy to understand and program.
\linebreak
\noindent
{\bf LiLAC-How} specifies how libraries can be used to perform a
LiLAC-What-specified computation.
Besides generating setup code and handling hardware context management, it 
crucially enables efficient memory synchronization.
It uses memory protection mechanisms to automatically track data changes and
transfers memory only when necessary.

The research contribution of this paper is a combination of three techniques for
the acceleration of sparse linear algebra:
\begin{itemize}
\item Accelerate unchanged source code by identifying sparse linear algebra
      computations with backtracking search.
\item Avoid vendor lock-in with an extensible specification language that adapts
      to new accelerator libraries.
\item Achieve program-wide memory synchronization with only local
      transformations using memory protection.
\end{itemize}
Together, these techniques result in a system that works on existing and novel
software.
It offers the full performance of fast libraries, avoids vendor lock-in, and
keeps the source code easy to maintain and free from pollution.

\section{Overview}
\label{sec:overview}

\autoref{motivationcode} shows the LiLAC-enabled compiler from the user
perspective.
In the top left corner (1), we see unmodified application source code.
This is {\em conjugate gradient} from the NAS-PB suite.
To achieve good performance on Intel processors, the compiler (2)
has been configured to offload native sparse code to Intel MKL.
Using a specification of {\em What} computations MKL supports, it
recognizes the highlighted loop as a suitable sparse matrix-vector product.
Instead of passing it on to the compiler backend for code generation,
it inserts a call to a {\em harness} function.
This is performed on intermediate code (3) and results in a program (4).
In the bottom left (5) is an equivalent source-level representation.

LiLAC also generates the corresponding harness code (6), which gets compiled
into a shared library (7) that is linked with the application binary.
This harness interfaces with the underlying library implementation,
Intel MKL (8).

\begin{figure*}[t]
\includegraphics[width=\textwidth]{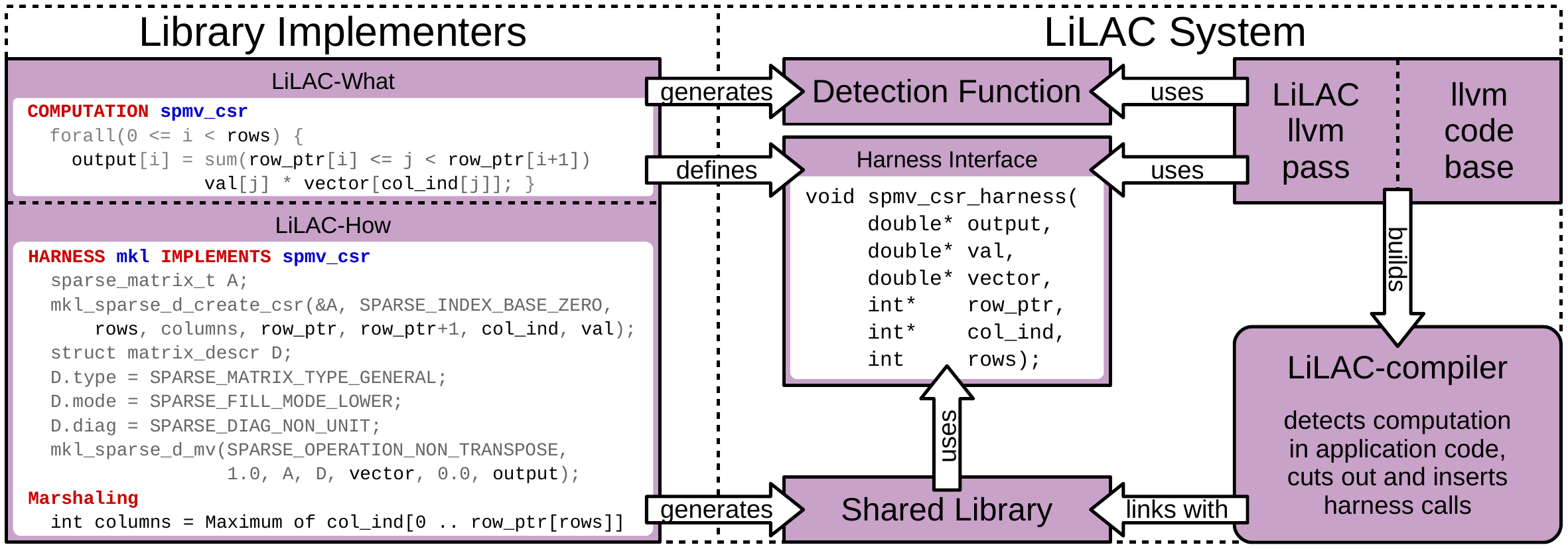}\\[-0.5em]
\caption{Overview of LiLAC internals: On the left is the complete LiLAC
         program that the library implementer has to provide.
         At compile time of LLVM, this program is parsed and incorporated
         into a modified clang C++ compiler, behaving as in
         \autoref{motivationcode}.\parfillskip=0pt}
\label{motivationflow}
\end{figure*}

\subsection{Implementation Overview}

\autoref{motivationflow} shows the internals of the LiLAC system.
It is fully integrated into the build system of the established LLVM compiler
framework, extending the clang compiler.

On the left is the LiLAC specification - just 16 lines of code.
It is independent of the user application and can be provided by the library
implementer.
It consists of a {\em What} and a {\em How} part.
These two parts are processed by the LiLAC system and result in a runtime
library and a generated detection function, which is incorporated into the
clang compiler.

{\bf LiLAC-What} specifies the functionality that is provided by a library,
in this example {\em spmv-csr} (cf.\ \autoref{motivationflow}).
From this, a function that detects the computation in normalized LLVM IR
code is generated and the harness interface
is determined.
The detection functions are based on a backtracking search algorithm, as
elaborated in \autoref{sec:implementation}.
The detection function is linked directly into the LiLAC-compiler, either
statically or dynamically at (compiler) run time.

{\bf LiLAC-How} specifies how the library, Intel MKL in this case, is invoked to
perform the specified calculation.
This involves boilerplate code, but also advanced features.
These include efficient data synchronization and the caching of invariants.
In the given example, the {\em columns} variable is such an invariant.
It is required for the library call, but not statically available.
Therefore, it has to be computed at runtime.
Using {\em Marshaling}, LiLAC automatically generates the harness such that
this is only recomputed if the values in {\em row\_ptr} change.
Such changes are captured with generated memory protection code using
{\em mprotect}, managed by LiLAC.

On the right of the figure, we can see how the components generated from the
LiLAC specification are used to build the LiLAC-compiler.
The detection function is compiled and used directly by the LiLAC-Compiler,
linked either statically or dynamically.
Interacting with the internals of LLVM, it implements a transformation
pass that is executed after the normal optimization pipeline.
Using the generated detection function, it finds instances of the computation
and replaces them with calls to the specified harness interface.

The harness, on the other hand, is compiled into a shared library.
The LiLAC-compiler dynamically links applications to this shared library
whenever it inserts harness calls.
When multiple LiLAC-How programs are provided, the generated harnesses
are compatible and linking the user program to a different harness library at
runtime is sufficient.

\section{What and How}
\label{sec:lilacwhat}
This section describes in more detail the two components of the LiLAC language.
LiLAC-What specifies the computations that a library performs;
LiLAC-How describes how exactly the library should be invoked to perform these
computations.

\begin{figure}[p]
\input{figure_lilacwhat}\\[-0.5em]
\caption{Grammar of the LiLAC-What language}
\label{bnfgrammar}
\vspace{1.5em}
\begin{minipage}[b]{0.3\linewidth}
\includegraphics[width=0.9\linewidth]{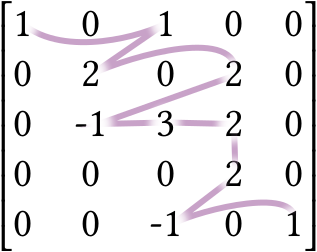}
\vspace{0.04em}
\end{minipage}
\begin{minipage}[b]{0.65\linewidth}
\begin{align*}
\text{\bf val} =& \begin{bmatrix}
1\ \ 1\ \ 2\ \ 2\ \ \text{-}1\ \ 3\ \ 2\ \ 2\ \ \text{-}1\ \ 1\\
\end{bmatrix}\\
\text{\bf col\_ind} =& \begin{bmatrix}
0\ \ 2\ \ 1\ \ 3\ \ 1\ \ 2\ \ 3\ \ 3\ \ 2\ \ 4\\
\end{bmatrix}\\
\text{\bf row\_ptr} =& \begin{bmatrix}
0\ \ 2\ \ 4\ \ 7\ \ 8\ \ 10\\
\end{bmatrix}
\end{align*}
\end{minipage}
\caption{Compressed Sparse Row (CSR) representation as used by the LiLAC-What example
         in \autoref{motivationcode} and \autoref{motivationflow}
         \parfillskip=0pt}
\label{csr_lilacwhat_fig}
\vspace{1.5em}
\centering
\begin{minipage}[b]{0.3\linewidth}
\includegraphics[width=0.9\linewidth]{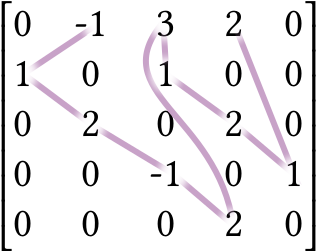}
\vspace{0.04em}
\end{minipage}
\begin{minipage}[b]{0.65\linewidth}
\footnotesize
\begin{align*}
\text{\bf perm} =& \begin{bmatrix}1\ \ 2\ \ 0\ \ 4\ \ 3\\\end{bmatrix}\\[-0.4em]
\text{\bf val} =& \begin{bmatrix}\text{-}1\ \ 1\ \ 2\ \ \text{-}1\ \ 2\ \ 3\ \ 1\ \ 2\ \ 1\ \ 2\\\end{bmatrix}\\[-0.4em]
\text{\bf col\_ind} =& \begin{bmatrix}1\ \ 0\ \ 1\ \ 2\ \ 3\ \ 2\ \ 2\ \ 3\ \ 4\ \ 3\\\end{bmatrix}\\[-0.4em]
\text{\bf jd\_ptr} =& \begin{bmatrix}0\ \ 5\ \ 9\ \ 10\\\end{bmatrix}\\[-0.4em]
\text{\bf nzcnt} =& \begin{bmatrix}3\ \ 2\ \ 2\ \ 2\ \ 1\end{bmatrix}
\end{align*}
\end{minipage}
\vspace{0.5em}
\hrule
\vspace{0.3em}
\includegraphics[width=\linewidth]{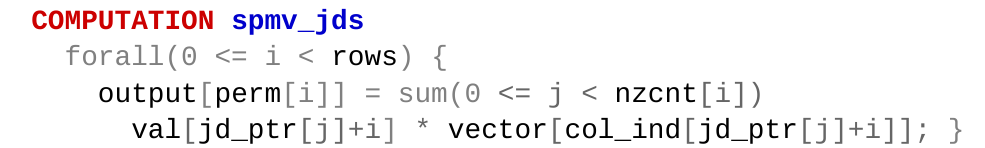}
\caption{Jagged Diagonal Storage (JDS) in LiLAC-What}
\label{jds_lilacwhat_fig}
\end{figure}

\subsection{LiLAC-What: Functional Description}
At the heart of our approach is a simple language to specify sparse and dense
linear algebra operations.
This serves two purposes in our LiLAC system: Firstly, it is used to generate
a detection program for finding the computation in user code.
Secondly, it identifies the variables that are arguments to the library, thus
defining the harness interface.

The key design challenge was to stay simple enough to automatically generate
robust detection functionality, yet to be able to capture operations in all
relevant data formats.
Most importantly, this includes the CSR/CSC, JDS and COO formats.
CSR and JDS are part of our evaluation.
Across the different formats, the control flow is rigid and easy to express.
This is reflected in the grammar as shown in \autoref{bnfgrammar}.

\subsection{Sparse Matrix Variations in LiLAC-What}
Sparse matrices can be stored in different formats.
We introduce two of them explicitly, but others are supported in the same way by
LiLAC-What.

\paragraph{Compressed Sparse Row (CSR) \cite{doi:10.1137/1.9780898718003}}
All non-zero entries are stored in a flat array \textbf{val}.
The \textbf{col\_ind} array stores the column position for each value.
Finally, the \textbf{row\_ptr} array stores the beginning of each row of the
matrix as an offset into the other two arrays.
The number of rows in the matrix is given directly by the length of the
\textbf{row\_ptr} array minus one, however, the number of columns is not
explicitly stored.
In \autoref{csr_lilacwhat_fig}, a 5x5 matrix is shown represented in this
format, the LiLAC-What code is in the top left of \autoref{motivationflow}.

\paragraph{Jagged Diagonal Storage (JDS) \cite{doi:10.1137/0910073}}
The matrix rows are reordered such that the number of non-zeros per row is
decreasing.
The permutation is stored in a vector \textbf{perm}, the number of nonzeros in
\textbf{nzcnt}.
The nonzero entries are then stored in an array \textbf{val} in the following
order: The first nonzero entry in each row, then the second nonzero entry in
each row etc.
The array \textbf{col\_ind} stores the column for each of the values and
\textbf{jd\_ptr} stores offsets into \textbf{val} and \textbf{col\_idx}.
The product of a sparse matrix in JDS format with a dense vector is specified 
in LiLAC-What at the bottom of \autoref{jds_lilacwhat_fig}.

\paragraph{Dense} Detecting dense is easier than sparse, and existing literature
covers it well.
We fully support dense but evaluate it only briefly for completeness.

\subsection{LiLAC-How}
\label{sec:lilachow}
Where LiLAC-What specifies the computations implemented by a library, LiLAC-How
describes how precisely library calls can be used to perform them.
The language was designed to support important existing libraries such as
cuSPARSE, clBLAS, and Intel MKL.
The idiosyncrasies of these libraries require LiLAC-How to capture some
boilerplate C++ code that manages the construction of parameter structures,
calling conventions etc.
Aside from this aspect, we designed it as high-level as possible without
compromising performance.
In particular, LiLAC-How abstracts away memory transfers.

These considerations result in two interacting components.
Firstly, a {\em harness} describes the boilerplate code for individual library
invocations.
Secondly, data {\em marshaling} between the core program and the
library is specified, which is crucial for heterogeneous compute
environments.
\autoref{lilacbnf2} shows the grammar specification of LiLAC-How.

\subsubsection{Individual Library Invocations}
We need to encapsulate the boilerplate code that any given library requires,
such as setup code, filling of parameter structures etc.
This part of the language is straightforward.

\begin{figure}[t]
\centering
\input{figure_lilachow}
\vspace{-1.5em}
\caption{Grammar of LiLAC-How}
\label{lilacbnf2}
\end{figure}

\paragraph{Harness}
The harness construct is the central way of telling the LiLAC system how a
library can be used to perform a computation that was specified in LiLAC-What.
As we can see at the top of \autoref{lilacbnf2}, a harness refers to a
LiLAC-What program by name and also has a name itself.
It is built around some C++ code, which can use all the variables from the
LiLAC-What program to connect with the surrounding program.
It also needs to specify the relevant C++ header files
that the underlying library requires.
Lastly, the harness can incorporate persistent state and utilize data
marshaling.

\paragraph{Persistence}
Many libraries need setup and cleanup code, which is specified with the
keywords {\em BeforeFirstExecution} and {\em AfterLastExecution}.
These are used in combination with {\em PersistentVariables}, allowing state to
persist between harness invocations, e.g.\ to retain handlers to hardware
accelerators.

\begin{figure}[p]
  \includegraphics[width=0.96\linewidth]{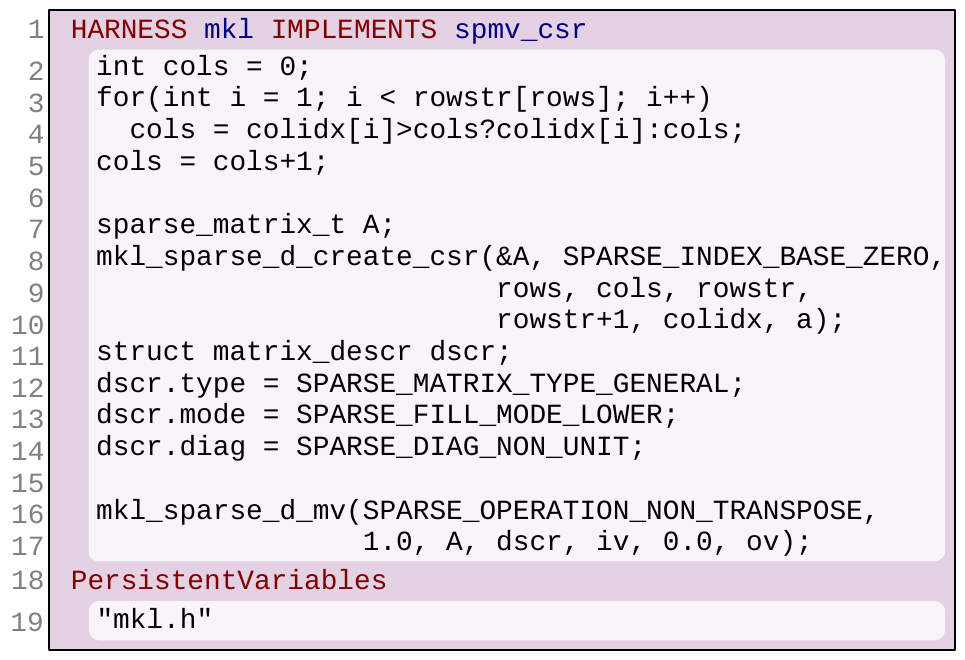}
\\[-0.75em]
\caption{This LiLAC-What program implements spmv-csr na\"ively with Intel MKL.
         Performance is degraded because of lines 2--5.
         \autoref{readablemax} will present a solution to this bottleneck.
         \parfillskip=0pt}
\label{mklharness1}
\vspace{0.75em}
\includegraphics[width=0.96\linewidth]{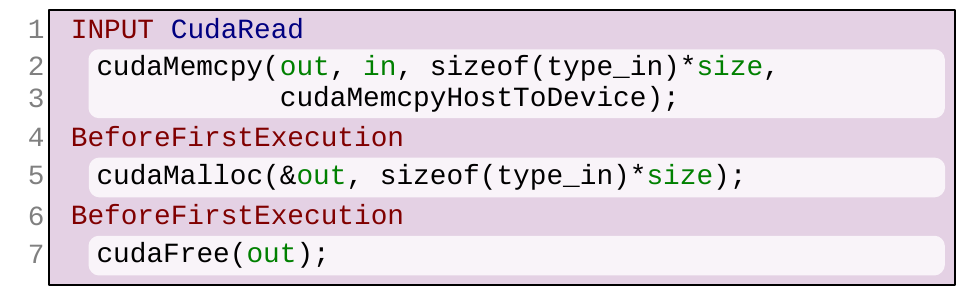}
\\[-0.75em]
\caption{LiLAC-How code to provide efficient automatic data marshaling between
         the host and the CUDA accelerator.\parfillskip=0pt}
\label{cudaread}
\vspace{0.75em}
\includegraphics[width=0.96\linewidth]{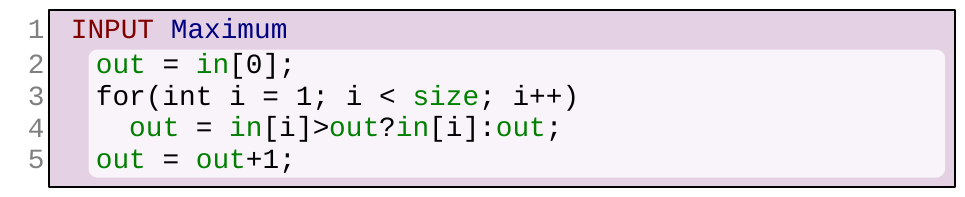}
\\[-0.75em]
\caption{{\em INPUT} can also be used to specify data-dependent computations
         that are only recalculated when necessary.\parfillskip=0pt}
\label{readablemax}
\vspace{0.75em}
\includegraphics[width=0.96\linewidth]{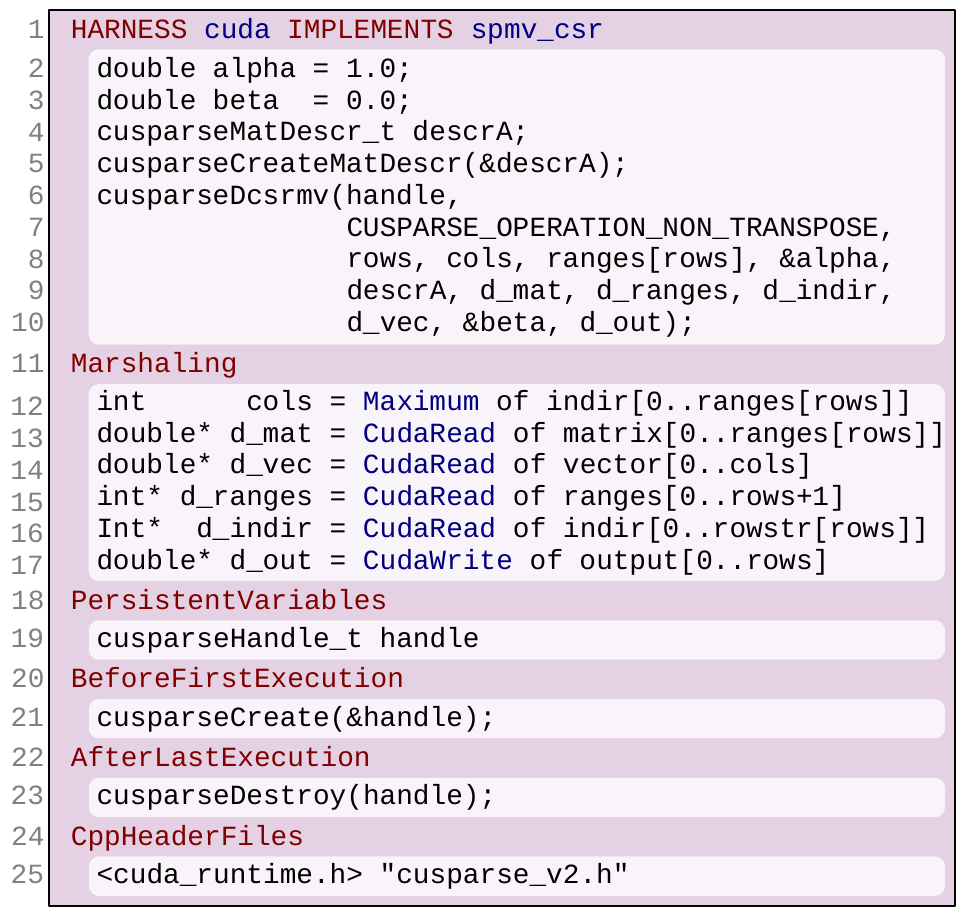}
\\[-0.75em]
\caption{This LiLAC-What specification implements an efficient SPMV harness
  using cuSPARSE in 25 lines of code.\parfillskip=0pt}
\label{spmvharness}
\end{figure}

\paragraph{Example}
In \autoref{mklharness1}, we see a trivial LiLAC-What program for implementing
\texttt{spmv\_csr} with the Intel MKL library.
The actual call to the relevant library function is in line 16.
To prepare for that call, there is boilerplate code in lines 7--14 to fill
parameter structures.

Critically, there is an additional parameter required by the library that is
data-dependent: the number of columns, {\em cols}, in the sparse matrix.
It is determined at runtime, in lines 2--5, leading to reduced performance.
We will avoid this with the data marshaling constructs in the next section.

\subsubsection{Data Marshaling}
Heterogeneous accelerators require data transfers to keep memory consistent
between host device and accelerator.
To achieve the best performance, these have to be minimized.

Importantly, unchanged data should never be copied again.
This requires program-wide analysis that is not available statically.
LiLAC-How uses memory protection to implement this at runtime with minimal
overhead by capturing read and write accesses to memory ranges.
The same mechanism is used to cache data-dependent invariants across several
invocations, such as {\em cols} in \autoref{mklharness1}.

Data {\em marshaling} routines are bound to ranges of memory in the harness.
In the specification, the underlying array is available using the
identifiers {\tt in}, {\tt size}, and {\tt out}.

\subsubsection{Detailed Example}
In \autoref{cudaread}, the \texttt{cudaMemcpy} function from NVIDIA CUDA is
integrated with LiLAC-How.
It is used to copy data from the host to the accelerator.
For this to work, it first needs to allocate memory of the
device using \texttt{cudaMalloc}, which is later freed with \texttt{cudaFree}.
Minimal memory transfers are obtained by executing \texttt{cudaMemcpy} only when
a value in the array changes.

We can use the same construct to efficiently compute values
such as the \texttt{cols} variable in \autoref{mklharness1}, as shown in
\autoref{readablemax}.
The optimized implementation is derived from \autoref{mklharness1} lines 2-5.
However, instead of the concrete variable
names, the reserved identifiers {\tt in}, {\tt size}, and {\tt out} are used.

\autoref{spmvharness} shows an \texttt{spmv\_csr} LiLAC-How program for the
cuSPARSE library.
A number of data marshaling variables are introduced in lines 12--17, that
automatically optimize both memory transfers and the computation of the
\texttt{cols} variable.
The core of the harness in lines 2--10 is again nothing more than 
library-specific boilerplate C++ code.

\section{Implementation}
\label{sec:implementation}
The LiLAC system, as shown in \autoref{motivationflow} is entirely integrated
into the LLVM build system.
When LLVM is compiled, the LiLAC specification is parsed using a Python
program.
Based on the LiLAC-What and LiLAC-How sections, C++ code is generated that is
automatically incorporated into LLVM in further stages of the build process.

The result is an LLVM optimization pass that is available when linking LLVM
with the clang C/C++ compiler.
This pass performs the discovery of linear algebra code and the insertion of
harness calls.
Furthermore, the harness libraries themselves are built at compile time of LLVM,
using C++ code emitted from the LiLAC-How sections.

The two crucial implementation details are therefore the following:
Firstly, how automatic detection functionality in C++ is generated from the
LiLAC-What specifications.\linebreak
Secondly, how the LiLAC-How sections are used to generate fast C++
implementations of the specified library harnesses.

\subsection{LiLAC-What}
The parsed LiLAC-What sections are turned into C++ functions that recognize
places for harness call insertions in an LLVM pass.
This builds on previous work via a formulation in CAnDL
\cite{Ginsbach:2018:CDS:3178372.3179515}.
Detection is done on optimized compiler intermediate representation.
Standard \texttt{-O2} optimizations, excluding loop unrolling and
vectorization, normalize the intermediate code.
Optimizations minimize programming language-specific artifacts and the
impact of syntax-level programmer decisions.

The effect is demonstrated in \autoref{robustness}, which shows three
implementations of a dot product in different languages: C, C++, and FORTRAN.
After translating to LLVM IR and performing optimizations, the dot product is
recognized in the LiLAC system using the same LiLAC-What specification.

The detection comprises two steps, as demonstrated in \autoref{backtrack}.
Firstly, the control flow skeleton is recognized.
This is simple, as LiLAC-What can only express control flow in the form of loop
nests of a certain depth.
After candidate loop nests have been identified, the index and loop range
calculations from LiLAC-What are mapped onto the LLVM IR nodes.
This is done via a backtracking search procedure and allows robust detection
across many syntactically different input programs, as described in
\cite{Ginsbach:2018:CDS:3178372.3179515,Ginsbach:2018:AML:3173162.3173182}.

\subsubsection{Backtracking Search Algorithm}
For detecting instances of LiLAC-What specifications in user programs,
LLVM IR segments that match the control flow skeleton are identified.
These control flow candidates are then processed with a backtracking search
algorithm.

All $\left<exp\right>$ expressions in the LiLAC-What program are identified.
These have to be assigned instructions or other values from the LLVM IR segment.
Those top-level $\left<exp\right>$ expressions that are used as limits or
iterators in $\left<range\right>$ expressions are easily connected with the
corresponding loop boundaries in the control flow candidates.

The remaining expressions are successively assigned by backtracking.
Consider the example in \autoref{llvmexample}, which shows a candidate loop from
the LLVM IR generated from the C++ dot product code in \autoref{robustness}.
The iteration space is determined by loop analysis and this immediately allows us
to assign the iterator and range in \autoref{backtrack} on the left.
The LLVM IR values that correspond to \texttt{a[i]}, \texttt{a},
\texttt{b[i]}, \texttt{b}, \texttt{a[i]*b[i]} and \texttt{result} are then
searched for.
When a partial solution fails, the algorithm backtracks.
This happens in the example once, when no suitable multiplication can be found
in step 5.
If no complete solution can be determined, the control flow candidate is
discarded.

\begin{figure}[t]
\begin{lstlisting}[language=LiLAC,numbers=none]
COMPUTATION dotproduct
    result = sum(0 <= i < length) a[i] * b[i];
\end{lstlisting}
\begin{lstlisting}[language=C,numbers=none]
int i = 0;
while(i < N) {
    x += (*(A+i))*(*(B+i));
    i+=1; }
\end{lstlisting}
\vspace{-0.75em}
\begin{lstlisting}[language=C++,numbers=none]
for(int i = 0; i < vec_a.size(); i++)
    x += vec_a[i]*vec_b[i];
\end{lstlisting}
\vspace{-0.75em}
\begin{lstlisting}[language=FORTRAN,numbers=none]
DO I = 1, N, 1
    X = X + A(i)*B(i)
END DO
\end{lstlisting}
\vspace{-1em}
\caption{Syntactically different computations in C, C++, or FORTRAN are captured
         by one LiLAC-What specification.\parfillskip=0pt}
\label{robustness}
\vspace{0.25em}
\begin{lstlisting}[language=llvm,numbers=none]
; <label>:17:
  %18 = phi i64 [ 0, %10 ], [ %26, %17 ]
  %19 = phi double [ 0.0, %10 ], [ %25, %17 ]
  %20 = getelementptr double, double* %9, i64 %18
  %21 = load double, double* %20
  %22 = getelementptr double, double* %12
  %23 = load double, double* %22
  %24 = fmul double %21, %23
  %25 = fadd double %19, %24
  %26 = add nuw i64 %18, 1
  %27 = icmp ugt i64 %14, %26
  br i1 %27, label %17, label %15
\end{lstlisting}
\vspace{-1em}
\caption{LiLAC intercepts LLVM IR after optimizations.
         This ensures normalized and language-independent features.
         \parfillskip=0pt}
\label{llvmexample}
\vspace{1em}
\begin{tabular}{rcr|rclrlr}
       &              &      & a[i]        & $\leftarrow$ & {\small\bf 1:} & \hspace{-3.5mm} \%21                   &                &                     \\
       &              &      & a           & $\leftarrow$ & {\small\bf 2:} & \hspace{-3.5mm} \%9                    &                &                     \\
     i & $\leftarrow$ & \%18 & b[i]        & $\leftarrow$ & {\small\bf 3:} & \hspace{-3.5mm} \%21                   & {\small\bf 6:} & \hspace{-3.5mm} \%23\\
length & $\leftarrow$ & \%14 & b           & $\leftarrow$ & {\small\bf 4:} & \hspace{-3.5mm} \%9                    & {\small\bf 7:} & \hspace{-3.5mm} \%12\\
       &              &      & a[i] * b[i] & $\leftarrow$ & {\small\bf 5:} & \hspace{-3.5mm} \textcolor{red}{fail!} & {\small\bf 8:} & \hspace{-3.5mm} \%24\\
       &              &      & result      & $\leftarrow$ &                &                                        & {\small\bf 9:} & \hspace{-3.5mm} \%25\\
\end{tabular}
\vspace{-0.5em}
\caption{After finding a candidate loop and receiving some variables from loop
         analysis (left), the backtracking solver attempts to assign the
         remaining variables one by one (right).\parfillskip=0pt}
\label{backtrack}
\end{figure}

\subsubsection{Code Replacement}
Each loop nest that matches a LiLAC-What specification is replaced with a
harness call.
To minimize the invasiveness of our pass, this is performed as follows:
Firstly, a harness call is inserted directly before the loop.
The function call arguments are selected from the backtracking result and passed
to the harness.
Secondly, the LLVM instruction that stores the result of the computation or
passes it out of the loop as a phi node is removed.
The remainder of the loop nest is removed automatically by dead code
elimination.

\begin{figure}[t]
\begin{lstlisting}
template<typename type_in, typename type_out>
void CudaRead_update(type_in* in, int size,
                     type_out& out) {
    cudaMemcpy(out, in, sizeof(type_in)*size,
               cudaMemcpyHostToDevice);
}
template<typename type_in, typename type_out>
void CudaRead_construct(int size, type_out& out) {
    cudaMalloc(&out, sizeof(type_in)*size);
}
template<typename type_in, typename type_out>
void CudaRead_destruct(int size, type_out& out) {
    cudaFree(out);
}
template<typename type_in, typename type_out>
using CudaRead = ReadObject<type_in, type_out,
    CudaRead_update<type_in,type_out>,
    CudaRead_construct<type_in,type_out>,
    CudaRead_destruct<type_in,type_out>>;
\end{lstlisting}
\vspace{-0.75em}
\caption{LiLAC uses code from \autoref{cudaread} to define three functions that
         specialize the ReadObject template, which uses \texttt{mprotect} for
         capturing memory accesses internally.\parfillskip=0pt}
\label{templatecode}
\end{figure}

\subsection{LiLAC-How}
LiLAC-How syntax elements that take C++ code generate
generic functions, and template parameter deduction inserts concrete types
during the compilation process.

In \autoref{templatecode}, we see the correspondence between generated C++
template functions and the specification in \autoref{cudaread}.
The three function bodies are directly inserted.
The functions are used to specialize the {\em ReadObject} class template, which
guarantees the following properties via memory protection:
\linebreak
\noindent
{\bf {\em construct}} is called before the first invocation and when
\texttt{in} or \texttt{size} change for consecutive harness invocations.
\linebreak
\noindent
{\bf {\em update}} is called after {\em construct}
and if any of the data in the array is changed between
consecutive harness invocations.
\linebreak
\noindent
{\bf {\em destruct}} is called in between consecutive {\em construct}
calls and before the program terminates.



\subsection{FORTRAN}
The LLVM frontend for FORTRAN under active development, flang, is in an
unfinished state and produces unconventional LLVM IR code.
Significant additional work was required to normalize the IR code.
We developed normalization passes in LLVM to overcome the specific
shortcomings, enabling FORTRAN programs to be managed as easily as C/C++.

The problems that we encountered included: differing
indexing conventions requiring offsetting pointer variables 
on a byte granularity with untyped pointers;
incompatible intermediate representation types
where all parameters are passed in as \texttt{i64} pointers,
frequently necessitating a pointer type conversion followed
by a load from memory; obfuscated loops with additional induction variable that
counts down instead of up such that the standard LLVM \textbf{indvars} pass is
unable to merge the loop iterators.


\section{Experimental Setup}
\label{sec:experimentalsetup}

\begin{table}[t]
  \begin{tabularx}{\columnwidth}{cXl}
    \textbf{Name} & \textbf{Hardware} & \textbf{Libraries} \\
    \toprule
    \toprule
    Intel-0 & 2$\times$ Intel Xeon E5-2620 \newline Nvidia Tesla K20 GPU
           & \multirow{4}{*}[-0.3em]{
                \begin{tabular}{l} 
                  MKL \\ 
                  cuSPARSE \\
                  clSPARSE \\
                  SparseX
                \end{tabular}
              }\\[1.5em]
    Intel-1 & Intel Core i7-8700K \newline Nvidia GTX 1080 GPU 
           & \\[.8em]
    \midrule
    AMD & AMD A10-7850K \newline AMD Radeon R7 iGPU \newline Nvidia Titan X GPU 
         & \multirow{3}{*}{
              \begin{tabular}{l}
                cuSPARSE \\
                clSPARSE $\times 2$\\
                SparseX
              \end{tabular}
            }\\[1.8em]
  \end{tabularx}
\vspace{0.25em}
  \caption{Evaluated platforms and library harnesses;
           AMD-0 supports clSPARSE on both its internal and its external GPU.
           \parfillskip=0pt}
  \label{tab:hardware}
\vspace{-0.5cm}
\end{table}

We wrote short LiLAC programs for a collection of linear algebra libraries 
and applied our approach to a chemical simulation application, two graph
analytics applications and a collection of standard benchmark suites.

\vspace{0.25em}
\noindent
{\bf\em Libraries\ }
We selected four different libraries for sparse linear algebra functions.
These were: Intel MKL \cite{mkl}, Nvidia cuSPARSE \cite{cusparse},
clSPARSE \cite{clsparse} and SparseX \cite{sparsex}.
MKL is a general-purpose mathematical library, while clSPARSE and cuSPARSE are
OpenCL and CUDA implementations of sparse linear algebra designed to be executed
on the GPU, and SparseX uses an auto-tuning model and code generation to optimize
sparse operations on particular matrices.

\vspace{0.25em}
\noindent
{\bf\em Applications\ }
To evaluate the impact of LiLAC in a real-world context, we used the
\textsc{pathsample} physical chemistry simulation suite, a large FORTRAN legacy
application \cite{doi:10.1080/00268970210162691} consisting of over 40,000 lines
of code.
Recent work shows that applications in this area are amenable to acceleration
using sparse linear algebra techniques \cite{SUTHERLANDCASH2017288}, and
\textsc{pathsample} provides a useful example of this.
We also evaluated two modern C++ graph analytics kernels (BFS and PageRank
\cite{demetrescu2009shortest,Beamer2015GAP}).
\textsc{pathsample} was run in two different modes and three different levels of
pruning, in each case using a system of 38 atoms \cite{doi:10.1063/1.478595}
commonly used to evaluate applications in this domain.
The graph kernels were run against 10 matrices from the University of Florida's
sparse matrix collection \cite{Davis:2011:UFS:2049662.2049663}, with sizes
between 300K and 80M non-zero elements.

\vspace{0.25em}
\noindent
For completeness and validation that our LiLAC-generated implementations were
correct, we also applied our technique to sparse programs from standard benchmark
suites: CG from the NAS parallel benchmarks \cite{Bailey1991NPB}, spmv from
Parboil \cite{Stratton2018} and the Netlib sparse benchmark suites
\cite{Dongarra2001}.
Each benchmark suite was run using their supplied inputs.

\paragraph{Platforms}
We evaluated our approach across 3 different machines with varying hardware
performance and software availability. Each one was only compatible with a
subset of our LiLAC-generated implementations---a summary of these machines is
given in \autoref{tab:hardware}.


\section{Results}
\label{sec:results}
\begin{figure*}[t]
  \centering
  \includegraphics[width=1.7\columnwidth]{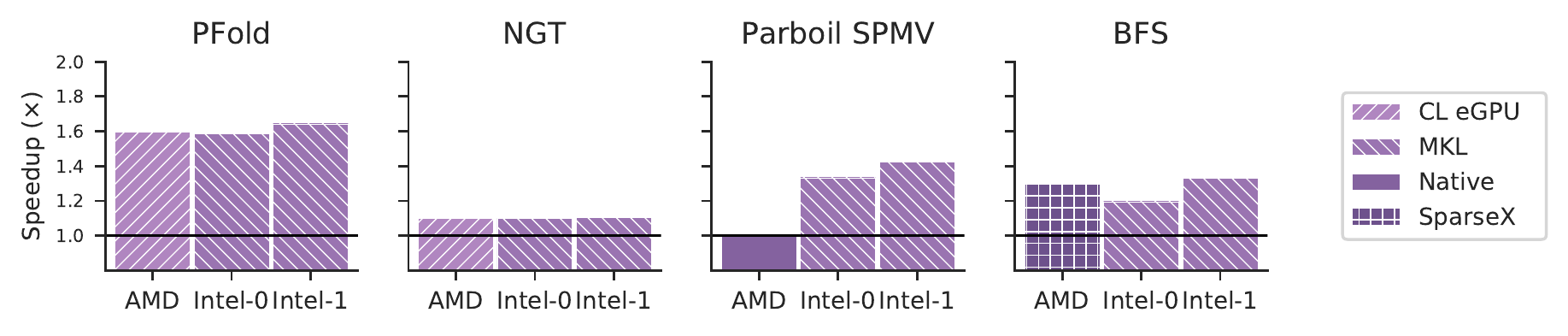}
  \includegraphics[width=1.7\columnwidth]{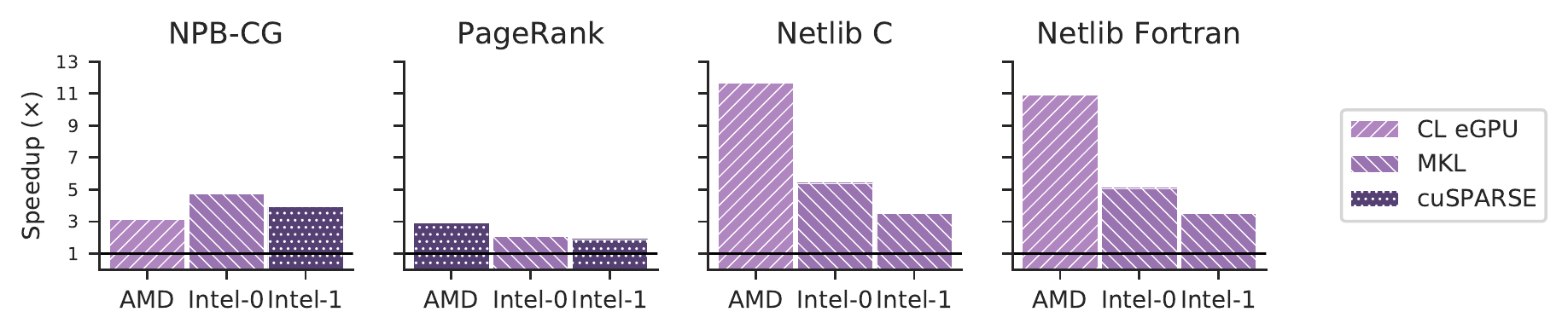}
\\[-1em]
  \caption{Evaluation on real-world applications and well-known benchmarks:
           Bars show the geomean speedup of the best-performing LiLAC harness
           across the set of input examples for each program and platform.
           Hatchings encode the selected implementations.
           The baseline is the identical source code compiled with clang
           \texttt{-O2}, yielding sequential CPU-only programs.
           \parfillskip=0pt}
\vspace{-0.5em}
  \label{spmv_app_perf}
\end{figure*}

We first present raw performance impact, then we analyze two intermediate
metrics: reliability of linear algebra discovery and effectiveness of memory
transfer optimizations.

\subsection{Performance}

LiLAC achieves significant speedups on real applications as well as benchmarks,
as shown in \autoref{spmv_app_perf}.
Baselines were compiled with \texttt{-O2} using the same version of clang
without LiLAC extensions.
Higher optimization levels (\texttt{-O3}) had a negligible impact on performance.
Different platforms and applications profit from different libraries
(\autoref{varietysection}).
Speedup ranges from $1.1$--$3\times$ on the scientific application codes to
12$\times$ on well-known sparse benchmark programs.

\paragraph{Applications}
On the \textsc{pathsample} applications (PFold and NGT), we measured
consistent speedups of approximately 50$\%$ and 10$\%$ respectively across
all 3 platforms.
For large applications, Amdahl's law is a severe limitation for
approaches like ours -- other parts of the applications dominate execution
times when linear algebra is accelerated.

\paragraph{Graph kernels}
PageRank requires a large number of SPMV calls using the same input matrix to
iterate until convergence.
The GPU implementations running on AMD and Intel-1 take advantage of
data remaining in memory.
The larger number of CPU cores and slower GPU available on \mbox{Intel-0} make
MKL its best-performing implementation.
CPU implementations perform best on BFS by avoiding memory copies entirely
-- on AMD, SparseX outperforms GPU implementations.

\paragraph{Benchmarks}
LiLAC achieves speedups of up to $12\times$ on 
standard sparse linear algebra benchmarks.
The impact is independent of the source language, as the C and FORTRAN versions
of the Netlib benchmark demonstrate.
LiLAC is able to achieve consistent, useful speedups across a variety of
hardware configurations.

\begin{table*}[t]
  \centering
  \small
  \begin{tabular}{ll|ccc|ccc|ccc|ccc}
    \multirow{2}{*}{\textbf{Platform}} &
    \multirow{2}{*}{\textbf{Implementation}} &
    \multicolumn{3}{c}{\textbf{PFold}} & \multicolumn{3}{c}{\textbf{NGT}} &
    \multicolumn{3}{c}{\textbf{PageRank}} & \multicolumn{3}{c}{\textbf{BFS}} \\[-0.2em]
    & & \hspace{0.5em}L0\hspace{0.5em} & L1 & \hspace{0.5em}L2\hspace{0.5em}
      & \hspace{0.5em}L0\hspace{0.5em} & L1 & \hspace{0.5em}L2\hspace{0.5em}
      & Erdos & \hspace{-0.5em}LJ-2008\hspace{-0.5em} & Road
      & Erdos & \hspace{-0.5em}LJ-2008\hspace{-0.5em} & Road \\
    \hline
    \hline
    \multirow{4}{*}{AMD}
    & cuSPARSE        & 1.38       & 1.18       & 0.67       & 0.69       & 0.69       & 0.70       & {\bf 3.44} & {\bf 1.18} & {\bf 9.97} & 1.62       & 6.55  & 1.96 \\[-0.3em]
    & clSPARSE (eGPU) & {\bf 2.17} & {\bf 1.82} & {\bf 1.22} & {\bf 1.16} & {\bf 1.16} & {\bf 1.16} & 3.08       & 1.24       & 6.06       & 0.50       & 11.03 & 0.24 \\[-0.3em]
    & clSPARSE (iGPU) & 2.03       & 1.78       & 1.03       & 0.90       & 0.90       & 0.90       & 3.26       & 1.31       & 4.05       & 0.14       & 4.17  & 0.05 \\[-0.3em]
    & SparseX         & -          & -          & -          & -          & -          & -          & -          & -          & -          & {\bf 1.93} & -     & - \\
    \hline
    \multirow{3}{*}{Intel-0} 
    & MKL      & {\bf 2.88} & {\bf 2.46} & {\bf 1.00} & {\bf 1.18} & {\bf 1.18} & {\bf 1.18} & {\bf 1.25} & {\bf 2.93} & {\bf 1.72} & {\bf 2.50} & {\bf 1.06} & {\bf 1.05} \\[-0.3em]
    & cuSPARSE & 0.75       & 0.60       & 0.45       & 0.66       & 0.66       & 0.66       & 1.39       & 1.00       & 3.32       & 0.87       & 1.74       & 1.28 \\[-0.3em]
    & clSPARSE & 0.90       & 0.75       & 0.46       & 0.81       & 0.79       & 0.78       & 1.24       & 0.95       & 2.24       & 0.13       & 1.45       & 0.07 \\[-0.3em]
    & SparseX & - & - & - & - & - & - & - & - & - & 1.19 & - & - \\
    \hline
    \multirow{3}{*}{Intel-1}
    & MKL      & {\bf 2.70} & {\bf 2.43} & {\bf 1.01} & {\bf 1.20} & {\bf 1.20} & {\bf 1.19} & 1.63       & 1.03       & 2.26       & {\bf 1.06} & {\bf 2.09} & {\bf 1.27} \\[-0.3em]
    & cuSPARSE & 0.48      & 0.41        & 0.30       & 0.68       & 0.69       & 0.68       & {\bf 1.59} & {\bf 0.87} & {\bf 4.44} & 1.01 & 1.83 & 1.63 \\[-0.3em]
    & clSPARSE & 1.00      & 1.00        & 1.00       & 1.00       & 1.02       & 1.00       & 1.50       & 0.87       & 3.46       & 0.23 & 1.81 & 0.13 \\[-0.3em]
    & SparseX & - & - & - & - & - & - & - & - & - & 1.25 & - & - 
  \end{tabular}
\vspace{0.25em}
  \caption{LiLAC speedups on each platform, across different applications and
  problem sizes.
  SparseX demonstrated promising performance on some applications, but we were
  unable to evaluate on every relevant instance due to instability.
  Implementation with best geomean speedup per benchmark and platform is bold.}
  \label{tab:perf_data}
\vspace{-2.8em}
\end{table*}

\paragraph{Dense}
We evaluated on some dense benchmarks as well.
In line with the literature, dense is very amenable to heterogeneous
acceleration.
We achieve $20\times$ speedup on Parboil sgemm by inserting LiLAC-harnessed
calls into sequential baseline.
However, impressive heterogeneous speedups on dense are well explored in the
literature, we focus on sparse.

\begin{figure}[t]
  \centering
  \includegraphics[width=0.75\columnwidth]{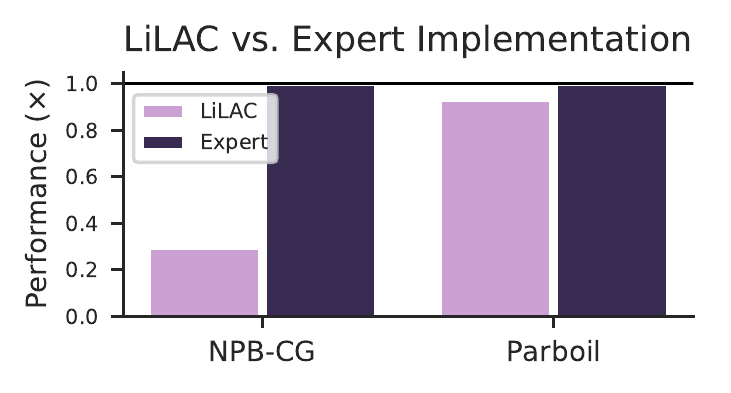}
  \begin{tabular}{lrr}
    \multirow{2}{*}{\textbf{Benchmark}} & \multicolumn{2}{c}{\textbf{Modified LoC}} \\
    & LiLAC & Expert \\
    \toprule
    NPB-CG       & 0 (44) & 1948 \\
    Parboil SPMV & 0 (44) & 261
  \end{tabular}
  \caption{LiLAC performance as fraction of expert version performance.
           We achieve good speedup with no application programmer effort
           (measured as required LoC change).
           The LiLAC required code -- identical across programs -- is in
           parentheses.
           Amdahl's Law limits our impact on NPB-CG.\parfillskip=0pt}
  \label{tab:lilac-linesofcode}
\end{figure}

\paragraph{Comparison to Expert}
NPB and Parboil contain expert-written alternative versions with GPU
acceleration.
This allowed the evaluation of LiLAC against heterogeneous code reaching close
to peak performance, shown in \autoref{tab:lilac-linesofcode}.

While the expert version of NPB-CG is $\sim 3\times$ faster, this is not due
to an improved sparse linear algebra operation, but a complete
parallelization and rewrite of the program for the GPU.
In Parboil SPMV, the expert version focuses on improved sparse linear algebra.
Here the difference between an expert and LiLAC is only 1.07$\times$.

\paragraph{Productivity}
The bottom of \autoref{tab:lilac-linesofcode} shows the amount of code modified
in order to add heterogeneous acceleration manually vs with LiLAC.
This demonstrates the productivity improvements for application programmers.
No lines of user code need to be modified using LiLAC, while both expert
versions require significant application rewrites.
Only 44 lines of application-independent LiLAC code is required.

\subsection{Necessity of Flexible Backends}
\label{varietysection}
\begin{figure}[t]
  \centering
  \includegraphics[width=\columnwidth]{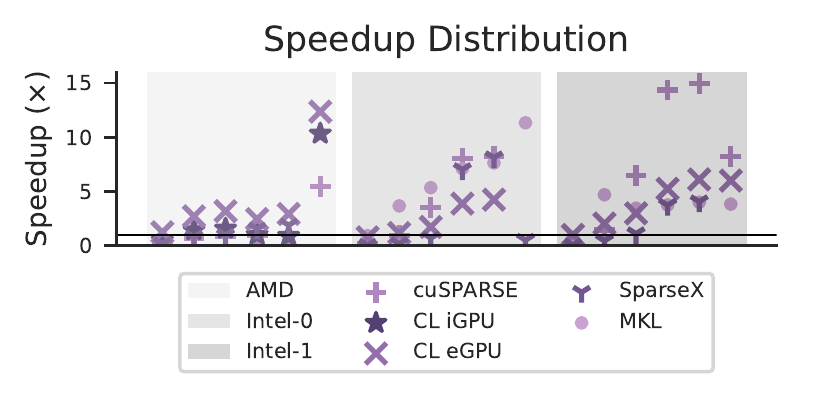}
\\[-1.7em]
  \caption{Distribution of speedups on NPB-CG.
    The stacks within each of the three columns are sorted by problem size, each
    point shows the speedup of a specific implementation.\parfillskip=0pt}
  \label{fig:performance-distribution}
\end{figure}

The relative  performance of different accelerator libraries is highly dependent
on the application, problem size, and platform, as
\autoref{fig:performance-distribution} shows.

\autoref{tab:perf_data} has more detailed data.
The best-performing implementation varies considerably, depending on
characteristics of the problem in question.
No accelerator library performs well reliably, each harness outperforms
any other harness on some combination of data set and platform.
For some small problem sizes, hardware acceleration is not profitable.
Those slowdowns are due to inherent overheads, not LiLAC.

\subsection{Effectiveness of Data Marshaling}
Our implementation of LiLAC relies on a non-trivial data marshaling system that
prevents redundant computations and memory transfers. We present performance
results that show the importance and effectiveness of this system.

\begin{figure}[t]
  \centering
  \includegraphics[width=\columnwidth]{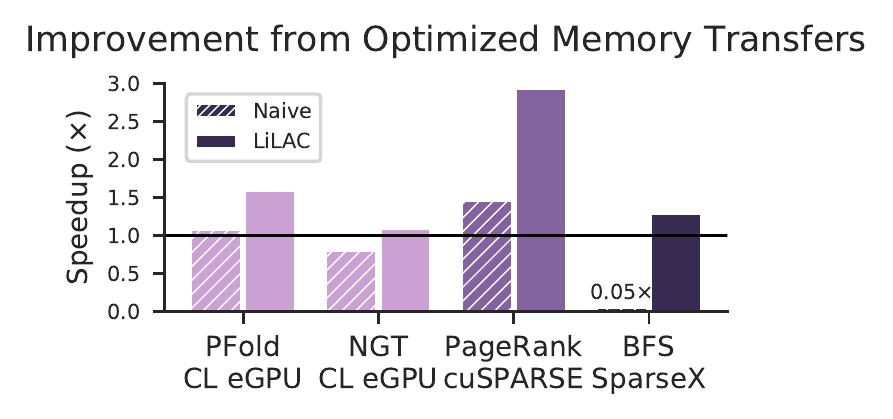}
\\[-1.3em]
  \caption{LiLAC vs. na\"ive library calls without marshaling optimizations,
           speedup over sequential baseline:
           Advanced marshaling features of LiLAC are critical for performance.
           \parfillskip=0pt}
  \label{fig:transfer-performance}
\end{figure}

We repeated our experiments, using the best-performing implementations from
\autoref{spmv_app_perf}.
Instead of using the data marshaling scheme, we recompute and transfer memory
naively for each invocation.
The results are in \autoref{fig:transfer-performance}.
Across the best AMD versions of PFold, NGT, PageRank and BFS -- where
accelerators are profitable with marshaling -- only PageRank achieves
a significant speedup naively.

For BFS, the naive approach leads to drastic performance degradation, the
marshaling version is $25\times$ faster.
This is because it performs an internal matrix tuning phase that is far more
expensive than a memory copy.
For the other three programs, there is a factor of 1.4--3.5$\times$ between
the naive and the smart version.


\begin{table}[b]
\vspace{-0.3cm}
\centering
\caption{Sparsity does not fit the polyhedral model;
         Polly is not available for FORTRAN;
         Intel compilers fail to parallelize sparse.
         Only LiLAC detects sparse linear algebra reliably.\parfillskip=0pt}
\label{compareiccpolly}
\vspace{-0.5em}
\small
\begin{tabular}{l|c|c|c}
{\bf Benchmark}      & {\bf LiLAC}    & {\bf Polly}   & {\bf Intel icc/ifort} \\
\hline
\hline
PFold          & CSR & - & parallel dependence \\[-0.3em]
NGT            & CSR & - & parallel dependence \\[-0.3em]
Parboil-SPMV   & JDS & no SCoP & parallel dependence \\[-0.3em]
BFS            & CSR & no SCoP & parallel dependence \\[-0.3em]
NPB-CG         & CSR & - & parallel dependence \\[-0.3em]
PageRank       & CSR & no SCoP & parallel dependence \\[-0.3em]
Netlib C       & CSR & no SCoP & parallel dependence \\[-0.3em]
Netlib Fortran & CSR & - & parallel dependence \\
\end{tabular}
\end{table}

\subsection{Reliability of Discovery}
For performance impact, LiLAC needs to first detect linear
algebra computations.
Previous results already implied that this works reliably, and
\autoref{compareiccpolly} \mbox{reiterates} this.
All relevant sparse matrix-vector multiplications were recognized.

Established approaches, like the polyhedral model, are unable to model sparse
linear algebra, as verified with the Polly compiler.
Similarly, the Intel C/C++ and FORTRAN compilers fail to auto-parallelize, as
they cannot reason about sparsity and have to assume additional dependencies.

These results show the novelty of the abilities of LiLAC rather than
implementation weaknesses of Polly and ICC, as neither were designed for
accelerating sparse computations.



\section{Related Work}
\label{sec:relatedwork}
\paragraph{Compiler centric linear algebra optimization}
Compiler management of indirect memory accesses was first examined using an
inspector-executor model for distributed-memory machines
\cite{Baxter:1989:RPS:72935.72967}.
The location of read data was discovered at runtime and appropriate
communication inserted.
Later work was focused on efficient runtime dependence analysis and the
parallelization of more general programs
\cite{pottenger1995idiom,fisher1994parallelizing,rauchwerger1999lrpd,
suganuma1996detection}.
However, the performance achieved is modest due to runtime 
overhead and falls well short of library performance.
More recent work developed equality constraints and subset relations that help
reduce the runtime overhead
\cite{Mohammadi:2019:SCD:3314221.3314646}.

The polyhedral model is an established compiler approach for modeling data
dependencies
\cite{redon1994scheduling,jouvelot1989unified,chi1997optimizing,
gupta2006simplifying,stock2014framework}.
Such an approach has been implemented in optimizing compilers, such as the Polly
extensions to LLVM \cite{Doerfert2015Polly}.
Recent work has extended the polyhedral model beyond affine programs to some
forms of sparsity with the PENCIL extensions \cite{7429301}.
These can be used to model important features of sparse linear algebra, such as
counted loops \cite{Zhao:2018:PCF:3178372.3179509}, i.e.\ loops with dynamic,
memory dependent bounds but statically known strides.
Such loops are central to sparse linear algebra.
The PPCG compiler \cite{Verdoolaege:2013:PPC:2400682.2400713} can detect
relevant code regions, but it relies on well behaved C code with all arrays
declared in variable-length C99 array syntax.
This excludes most real-world programs;
nothing in our evaluation fits this structure.

The Appollo system \cite{10.1145/2838734} integrates thread level speculation
with the polyhedral model, allowing its application to sparse linear algebra.
However, it requires sub-parts of the computation to perform dense accesses at
runtime.
Similar approaches \cite{10.1145/3314221.3314615} also require regular
sub-computations.

\paragraph{Compiler detection} Previous work has detected code structures
in compilers using constraint programming.
Early work was based on abstract computation graphs
\cite{pinter1994program}, but more recent approaches have used compiler
intermediate code and made connections to the polyhedral model
\cite{Ginsbach:2018:CDS:3178372.3179515}.

In \cite{Ginsbach:2018:AML:3173162.3173182} they implement a method that
operates on SSA intermediate representation.
It uses a general-purpose low-level constraint programming language aimed at
compiler engineers.
The paper focuses on code detection, with manual data marshaling.
Recent work \cite{pactpaper} uses type-guided program synthesis to
model library routines, which are then detected by a solver.
Again, data marshaling is not taken into account.

Other advanced approaches to extracting higher-level structures from assembly
and well-structured FORTRAN code involve temporal logic
\cite{Mendis2015Helium, Kamil2016Verified}.
These approaches tend to focus on a more restricted set of computations
(dense memory access).
While this allows formal reasoning about correctness, is too restrictive to
model sparse linear algebra.

\paragraph{Domain-Specific Languages}
There have been multiple domain-specific libraries proposed to formulate linear
algebra computations.
Many of these contain some degree of autotuning functionality to achieve good
performance across different platforms
\cite{Sujeeth:2014:DCA:2601432.2584665}.
Halide 
\cite{Ragan-Kelley:2013:HLC:2499370.2462176}
was designed for image processing.
\cite{Suriana:2017:PAR:3049832.3049863}.
Its core design decision is the scheduling model that allows the separation of
the computation schedule and the actual computation.
There has been work on automatically tuning the schedules
\cite{Mullapudi:2016:ASH:2897824.2925952} but in general, the computational
burden is put on the application programmer.


The SLinGen \cite{Spampinato:18} compiler takes a program expressed in the
custom LA language, inspired by standard mathematical notation.
It then implements custom code generation for the expressed calculations, with a
focus on small, fixed-size operands.
This is built on top of building blocks provided by previous work on LGen
\cite{Spampinato:14}.
The approach outperforms libraries focused on large data sizes but is
unable to utilize heterogeneous compute and requires program rewrites.

\paragraph{Libraries}
The most established way of encapsulating fast linear algebra routines is via
numeric libraries, generally based on the BLAS interface
\cite{2002:USB:567806.567807}.
These are generally very fast on specific hardware platforms, but require
application programmer effort and offer little performance portability.
Implementations of dense linear algebra are available for most suitable hardware
platforms, such as cuBLAS \cite{cublas} for NVIDIA GPUs, clBLAS \cite{clblas}
for AMD GPUs and the Intel MKL library \cite{mkl} for Intel CPUs and
accelerators.

Fast implementations of sparse linear algebra are fewer, but they exist for the
most important platforms, including cuSPARSE \cite{cusparse} and clSPARSE
\cite{clsparse}.
There have been several BLAS implementations that attempt platform independent
acceleration and heterogeneous compute \cite{Wang:2016:BHP:2925426.2926256,
10.1007/978-3-319-64203-1_33, Diego2017Multi}.


\paragraph{CPU-GPU data transfer optimizations}
Data transfers between CPU and GPU have been studied extensively as an important
bottleneck for parallelization efforts.
Previous work \cite{Jablin:2011:ACC:1993316.1993516,10.1145/2544137.2544156}
established systems for automatic management of CPU-GPU communication.
The authors of \cite{Lee:2009:OGC:1594835.1504194} implemented a system to
move OpenMP code to GPUs, optimizing data transfers using data flow analysis.
However, this approach performs a direct translation, not optimizing the code
for the specific performance characteristics of GPUs.

\section{Conclusion}
\label{sec:conclusion}



This paper presented LiLAC, a language and compiler that enables existing
codebases to exploit sparse (and dense) linear algebra accelerators.
No effort is required from the application programmer.
Instead, the library implementer provides a specification, which
LiLAC uses to automatically and efficiently match user code to
high-performance libraries.

We demonstrated this approach on C, C++, and FORTRAN benchmarks as well as
legacy applications, and shown significant performance improvement across
platforms and data sets.
In future work, we will investigate how our framework can be adapted to other
application domains, enabling effort-free access to an even larger set of
accelerator libraries.

\balance
\bibliography{references}{}

\end{document}